\documentclass[useAMS,usenatbib]{mn2e}

\usepackage{graphicx}

\title[Bias in C~IV-based quasar black hole mass scaling relationships from reverberation mapped samples]{Bias in
 CIV-based quasar black hole mass scaling relationships from reverberation mapped samples}
\author[M. S. Brotherton et al.]{Michael S. Brotherton$^{1}$\thanks{E-mail:
mbrother@uwyo.edu} , J. C. Runnoe$^{1,2}$, Zhaohui Shang$^{3}$, and M. A. DiPompeo
$^1$\\
$^{1}$Department of Physics and Astronomy, University of Wyoming, Laramie, WY 82071, USA\\
$^{2}$Department of Astronomy and Astrophysics, The Pennsylvania State University\\ 525 Davey Lab, University Park, PA 16803, USA\\
$^{3}$Department of Physics, Tianjin Normal University, Tianjin 300387, China}
\begin{document}		

\date{}

\pagerange{\pageref{firstpage}--\pageref{lastpage}} \pubyear{2013}

\maketitle

\label{firstpage}

\begin{abstract}
The masses of the black holes powering quasars represent a fundamental 
parameter of active galaxies. 
Estimates of quasar black hole masses using single-epoch spectra are quite 
uncertain, and require quantitative improvement.  
We recently identified a correction 
for C~IV $\lambda$1549-based scaling relationships used to estimate 
quasar black hole masses that relies on the continuum-subtracted 
peak flux ratio of the ultraviolet emission-line blend Si IV + OIV]
(the $\lambda$1400 feature) to that of C~IV.  This parameter correlates with
the suite of associated quasar spectral properties collectively known as 
``Eigenvector 1'' (EV1).
Here we use a sample of 85 quasars with quasi-simultaneous optical-ultraviolet 
spectrophotometry to demonstrate how biases in the average EV1 properties can 
create systematic biases in C~IV-based black hole mass scaling relationships.
This effect results in nearly an order of magnitude moving
from objects with small $<$ peak $\lambda$1400/C~IV $>$, which have overestimated
black hole masses, to objects with large $<$ peak $\lambda1400$/C~IV $>$, 
which have underestimated values.   
We show that existing reverberation-mapped samples of
quasars with ultraviolet spectra -- used to calibrate C~IV-based scaling
relationships -- have significant EV1 biases that result in predictions of
black hole masses nearly 50\% too high for the average quasar. 
We offer corrections and suggestions to account for this bias.  

\end{abstract}


\section{Introduction}

Black hole mass is a fundamental quasar property, and better measuring 
poorly known fundamental properties has the 
potential to lead to breakthroughs.  Moreover, 
correlations between the central black hole mass in galaxies, 
both active and inactive, and other properties such
as the stellar velocity dispersion and bulge luminosity (Marconi \& Hunt 2003;
Tremaine et al. 2002; Kormendy \& Ho 2013), 
may indicate a connection necessary to understanding galaxy 
evolution more generally. 

Several methods exist to directly measure or indirectly estimate the central 
black hole masses of quasars 
(see Shen 2013 for a review).  
A primary method of measuring black hole masses is via 
reverberation mapping (RM, e.g., Peterson 1993).
The underlying premise is that broad-line region (BLR) gas moves 
virially under gravity (Peterson \& Wandel 1999, Onken et al. 2003).  This means 
that BLR gas at a distance R$_{\rm BLR}$ with a velocity $\Delta$V
can be used to provide the central mass using an equation of the form:

\begin{equation}
M_{\rm BH} = f\frac{R_{\rm BLR} \Delta V^2}{G}
\end{equation}

The factor $f$\ includes our ignorance about geometry (and perhaps 
other complications), and is obtained by calibration against
inactive galaxies (e.g., Onken et al. 2004). 

RM programs provide the radius R$_{\rm BLR}$ via time lags 
between the continuum and line flux, while the velocity $\Delta$V is
measured from the variable portion of the emission-line profile. 
Both the Full Width at Half Maximum (FWHM) and the line dispersion 
$\sigma_{line}$ of the RMS profile have been used,
in conjunction with an appropriately calibrated $f$.  
We will refer to the quantity $R_{\rm BLR} \Delta V^2/G$ as the virial product.
Numerous campaigns over more than two decades 
(e.g. Clavel et al. 1991; Kaspi et al. 2000, 2007; Peterson et al. 2004, 
Bentz et al. 2009b; Denney et al. 2010) have provided virial products 
for over 50 objects, which are not necessarily representative
of active galaxies in general (e.g., Shen 2013).  

Because of the large observational resources required for RM, 
easier methods of estimating black hole masses have been developed, e.g., 
black hole mass scaling relationships (e.g., Vestergaard 2002).
Single-epoch (SE) spectra, rather than RMS spectra, provide line-profile 
measurements, while the radius-luminosity relationship 
(e.g., Kaspi et al. 2000; Bentz et al. 2006) provides $R_{\rm BLR}$ rather than a time lag.  
These two points permit the modification of equation 1 and lead to:  

\begin{equation}
M_{\rm BH} = f\frac{R_{\rm BLR} \Delta V^2}{G} = f\frac{\lambda L_{\lambda}^{\gamma} \Delta V^2}{G}.
\end{equation}

The exponent $\gamma$\ appears to be consistent with 0.5, 
at least when the host galaxy contribution to the continuum is accounted for 
(e.g., Bentz et al. 2009a), as expected for some simple BLR models 
(e.g., Netzer 1990).

In practice, the black hole mass can be calculated from
the combination of the unscaled mass $\mu$ and a constant
$a$, which provides the scaling factor:

\begin{equation}
$$\mu = \left(\frac{\Delta V}{{\rm 1000\ km\ s^{-1}}} \right)^2 
\left( \frac{\lambda L_{\lambda}}{\rm 10^{44}\ ergs\ s^{-1}} \right)^{\gamma}$$
\end{equation}

\begin{equation}
$${\rm log}\ M_{\rm BH} = {\rm log}\ \mu + a $$ 
\end{equation}

The scaling relationships are not very precise, unfortunately,
and may not be very accurate, either, as we shall demonstrate.
Black hole masses calculated with these equations are typically 
uncertain to factors
of 3-4 (Vestergaard \& Peterson 2006, hereafter VP06).
Given their utility, needing only a SE spectrum, however, 
it is desirable to improve existing relationships (in particular
for those using ultraviolet lines like C~IV $\lambda$1549, which is redshifted 
into the optical regime for high-z quasars).  
Despite possible biases in selection, RM samples have 
been used to calibrate the SE scaling relationships.  The formulations of 
VP06 and Vestergaard \& Osmer (2009) have been two often-employed examples.  
Recently Park et al. (2013), using more reliable reverberation masses and 
myriad small improvements, have updated the C IV-based scaling relationship.

There have been a number of concerns noted regarding black hole mass estimates
using C~IV (e.g., Croom 2011, Assef et al. 2011, Trakhtenbrot \& Netzer 2012, etc.).  
The primary criticism is that the FWHM of the C~IV line does not always 
correlate well with that of H$\beta$ in SE spectra, and that the former is 
often narrower than the latter, the reverse of the results for the RMS
profiles from RM campaigns (e.g., Onken et al. 2004).
The SE profile of C~IV probably does not reflect purely virial motions.
Denney (2012) compared the reverberating component of C~IV against average 
profiles, showing evidence for a low-velocity emission-line region that 
did not vary, consistent with low-velocity C~IV-emitting gas existing 
on much larger size scales (in accordance with analysis of gravitationally 
lensed systems, Sluse et al. 2011).  These non-varying line cores can be
interpreted as the ``intermediate-line region'' or ILR (Wills et al. 1993; 
Brotherton et al. 1994).  Denney also showed that a profile shape parameter 
($S = FWHM/\sigma_{line}$) measuring the contamination of the C~IV line by
this non-virial component correlates with the differences between 
C~IV and H$\beta$ mass estimates and could perhaps be used to correct 
C~IV-derived masses.
 
The degree of non-virial/ILR C~IV contamination also correlates 
with a suite of properties collectively known as ``Eigenvector 1''
(hereafter EV1; see Boroson and Green 1992; Brotherton \& Francis 1998; 
Sulentic 2007; etc.). In the ultraviolet part of the spectrum, 
these properties include the shape of the C~IV profile,
as well as certain emission-line ratios and the difference between peak 
velocities. Shen et al. (2008; see also Shen \& Liu 2012) find 
a bias in C~IV black hole masses associated with these emission-line velocity shifts.
Bian et al. (2012) also note a systematic difference
between C~IV and Mg II based black hole masses that related to 
the equivalent width of C~IV, another EV1 property. 

\begin{figure}
\begin{center}
\includegraphics[width=8.9 truecm]{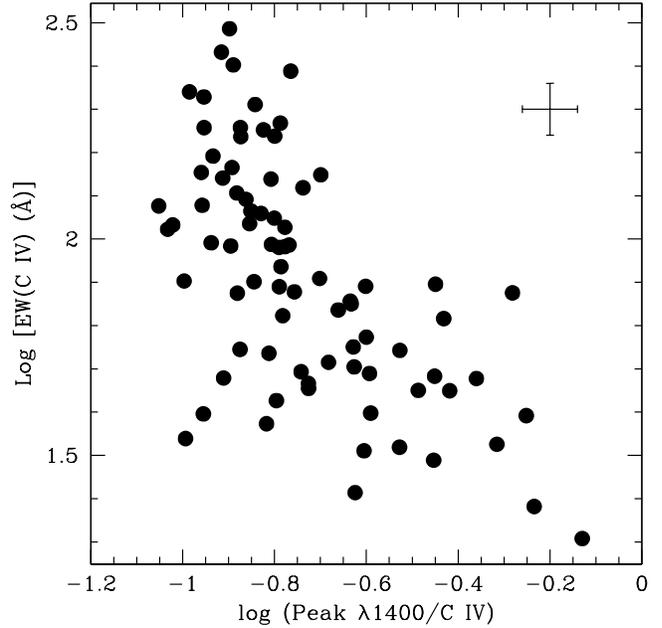}
\end{center}
\caption{
For the sample of Runnoe et al. (2013), we plot the log of
the continuum-subtracted peak ratio of the $\lambda$1400 feature versus the log
of the rest-frame equivalent width of C~IV $\lambda$1549.  This shows
that the peak ratio is also associated with EV1.
}
\label{fig:ew}
\end{figure}

Following up on Wills et al. (1993), Runnoe et al. (2013a, hereafter Paper 1)
developed a new C~IV mass correction relying on the log of the ratio of the 
continuum-subtracted peak heights of the $\lambda$1400 blend 
(of Si IV and O IV]) and C~IV.  Figure 1 shows a significant inverse correlation
between the peak ratio and the equivalent width of C~IV, demonstrating that this 
peak ratio is also part of EV1.   
Paper 1 provided a mass correction term based on this peak ratio that
improves the scatter between 
C~IV and H$\beta$ derived masses by $\sim$0.1 dex, or about 25\%.
There was also a suggestion of a bias in the VP06 C~IV 
scaling relationship, such that C~IV based masses systematically differed 
on average from H$\beta$, which could result in part from an EV1 bias in
the RM sample that it and other scaling relationships
are based upon.  Figure 2, showing distributions of a primary optical EV1 
parameter, peak $\lambda$5007, for the RM sample of Park et al.
(2013) and a complete sample of PG quasars  Boroson \& Green 1992), supports this 
hypothesis.   Quasars from Park et al. (2013) tend to have 
strong [O III] $\lambda$5007 emission, but rarely very weak 
[O III] $\lambda$5007 emission, compared to the representative, complete sample. 

\begin{figure}
\begin{center}
\includegraphics[width=8.9 truecm]{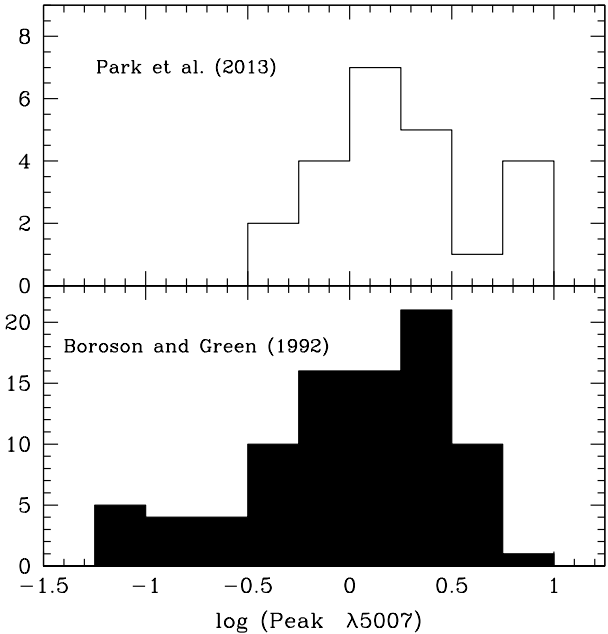}
\end{center}
\caption{
In the lower panel we plot the histogram of peak [O III] $\lambda$5007, 
a primary optical EV1 parameter, from the complete Palomar-Green subsample of 
Boroson \& Green (1992).
The top panel shows a histogram (based on estimates from literature
figures displaying spectra) for the same parameter for RM 
objects in the sample of Park et al. (2013), which has been used to 
calibrate SE C~IV black hole mass scaling relationships.  
}
\label{fig:o3hist}
\end{figure}

Here we will examine quantitatively 
how sample biases in EV1 lead to systematic shifts in estimates of 
black hole mass using C~IV-based scaling relationships, and the 
implications of this for determining what the best practices
should be going forward to pursue better black hole masses.
In particular we are concerned about commonly used 
scaling relationships based on RM samples that, 
while possessing superior and direct black
hole mass measurements, may be biased with respect to EV1. 

There may be several issues that affect SE C~IV black hole 
mass scaling relationships, and we use a careful approach to
isolate the EV1 bias.  It is beneficial to use consistent calibration,
measurements, and quasi-simultaneous optical-UV spectra when possible,
or risk offsets associated with those issues (e.g., see Denney et al. 2009 and
2013).  The presumably improved Park et al. (2013) equation
results in C~IV-based black hole masses systematically some 0.25 dex smaller
than those of VP06, for instance.

Ideally we would use the VP06 and Park et al. (2013) data and measurements,
but there are several problems with that approach.  First, they did not provide
the peak $\lambda$1400/CIV measurements we require to reproduce the 
Paper 1 analyses.  Second, after making our own measurements
as described in \S 2, the dispersion of peak $\lambda$1400/CIV 
in the RM objects is too small to significantly  correlate with the difference between
the C~IV and H$\beta$ line widths as it does in our paper 1 sample.  
For these reasons, we will create samples matched to the RM objects with our own data in order to determine
how the EV1 bias affects calculated masses.  By using our own high-quality 
data sets and our own measurements, we can avoid inconsistency.
 

In \S 3, we determine black hole mass scaling relationships for subsamples drawn
from the data set of Shang et al. (2011), each with different distributions of
peak $\lambda$1400/C~IV measurements from Paper 1, and demonstrate
the effect on SE scaling relationships.  We repeat the same
experiment for both nearly complete samples and for a carefully selected sample 
sample to show the effect of an EV1 bias on existing samples of RM quasars.
In \S 4, we discuss near-term and long-term resolutions to this
problem, which will allow C~IV to be used as a more effective mass
estimator.  Finally, \S 5 summarizes our conclusions.

\section{Data}

\subsection{SED Sample}

For our analysis, we use the sample from Paper 1, that of Shang et al.
(2011), which includes 85 quasars drawn from several sources.  
The unifying factor in assembling this composite sample was the existence of 
quasi-simultaneous spectrophotometry (within weeks or less), 
obtained in the near-ultraviolet taken with the 
{\em Hubble Space Telescope (HST)}, and in the optical from ground-based 
telescopes at Kitt peak National Observatory  and McDonald Observatory.  

One part was based on objects observed by the 
{\em Far Ultraviolet Spectroscopic Explorer, FUSE} (Moos et al. 2000), 
extending the spectra into the far-UV (see e.g. Shang et al. 2005).
Another subsample consists of radio-loud quasars specifically selected
to study orientation effects (see e.g., Wills et al. 1995, Runnoe et al. 2013b).
Finally, the last subsample is 22 of 23 objects in the complete
``PGX'' sample (Shang et al. 2007), which may be expected to be largely
unbiased with respect to quasar spectral properties.

Shang et al. (2011), Tang et al. (2012), and Paper 1 (Runnoe et al. 2013a)
provide measurements and determinations of many properties of the 
quasars in this sample, including peak $\lambda$1400/C~IV that is our ultraviolet EV1
indicator.  A valuable property of the sample is that it includes 
a large range of types of quasars with diverse properties.   
In particular, the complete 
PGX sample is useful to make sure parameter space is sufficiently covered and
suggestive of what the EV1 properties are of a representative quasar sample.

\begin{figure}
\begin{center}
\includegraphics[width=8.9 truecm]{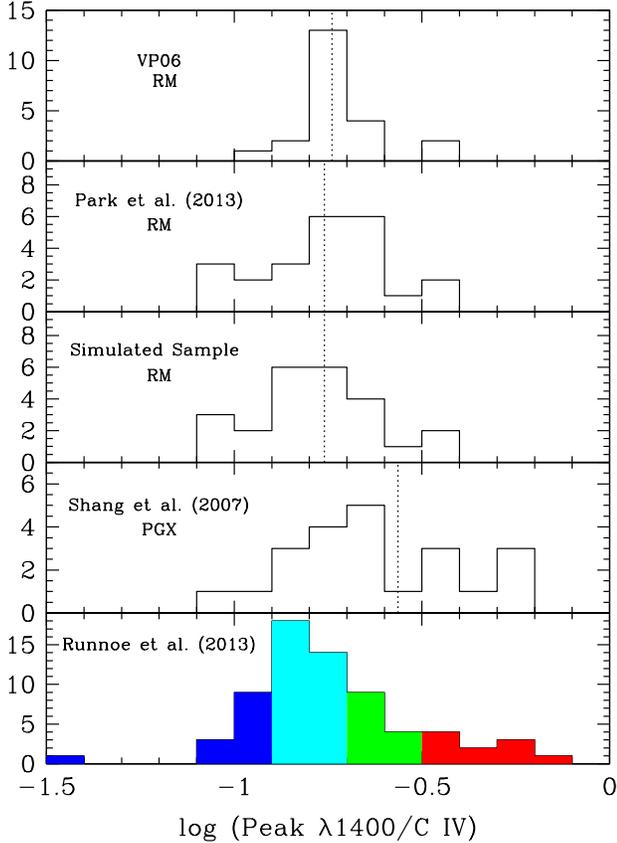}
\end{center}
\caption{Histograms of peak $\lambda$1400/C~IV for the samples used and
discussed in this paper, from bottom to top:
the sample of Runnoe et al. (2013), the PGX sample of Shang et al. (2007),
our subsample with peak $\lambda$1400/C~IV matched to the Park et al. (2013)
reverberation-mapped sample, and finally the reverberation-mapped samples
of Park et al. (2013) and VP06.  Dotted lines indicate the mean values
in these panels (from top to bottom, $-0.74$, $-0.76$, $-$0.77, and $-$0.56).  
Colors code the bottom histogram according to the log 
(peak $\lambda$1400/C~IV) ratio: blue ($<$0.9, cyan (from $-0.9$ to $-$0.7), 
green (from $-1$0.7 to $-$0.5), and red ($> -0.5)$.
We will use the same colors in later figures.
We draw attention to the PGX sample as being nearly complete, likely representing
an unbiased distribution of peak $\lambda$1400/C~IV and other properties.
}
\label{fig:hist}
\end{figure}

\subsection{Reverberation-Mapped (RM) Samples}

We want to know how possible biases in EV1 in heterogeneous 
RM samples may affect black hole masses obtained 
for the C~IV-based scaling relationships derived from them.
In order to do this, we need to know their peak $\lambda$1400/C~IV
distributions.


We use two RM samples that represent the basis 
for a commonly used C~IV-based black hole mass scaling relationship and
its recent update: VP06 and Park et al. (2013).  We obtained ultraviolet
spectra from the Multimission Archive at Space Telescope (MAST).
These come not only from {\em HST} spectrographs, but also 
the {\em Hopkins Ultraviolet Telescope (HUT),} and the {\em 
International Ultraviolet Explorer (IUE)}.  In a few cases where multiple
epochs of spectra were available for an object, we used the one that had the
best combination of signal-to-noise ratio and wavelength coverage
containing the $\lambda$1400 feature. 


We fit the $\lambda$1400 feature and C~IV line using a multiple Gaussian 
and a local continuum following the method of Tang et al. (2012) 
discussed in Paper 1.  Fits were individually inspected to ensure
that noise spikes and absorption features did not cause erroneous measurments.
We checked our fitting procedure against Paper 1
for the SED sample and found agreement within the uncertainties, as well as  
overall systematic agreement to better than a few per cent.
Our paper 1 approach was conservative, characterizing 
uncertainties on peak $\lambda$1400/C~IV as $\sim 15\%$.  
While this ensures measurement consistency between the 
samples, we noted that measurements of peak $\lambda$1400/C~IV for  
the same objects observed at different epochs usually agreed within 
15\% (e.g., Mrk 335 -- 0.16 vs. 0.15), but not always 
(e.g., NGC 7469 -- 0.17 vs. 0.10).  These differences are due to real,
intrinsic variability and serve as a caution to the use of non-simultaneous 
data. The peak $\lambda$1400/C~IV measurements span an order of magnitude, 
and that the observed variability is typically much smaller than that, 
but this is an issue of concern to investigate in the future.  


We also note that we do not include measurements for all objects in the samples
of  VP06 and Park et al. (2013).  From VP06, we have excluded NGC 3516,
NGC 4051, NGC 4151, and NGC 4593 because of absorption in C~IV making a peak
flux too difficult to measure reliably. 
Also excluded is PG 1307+085, for 
which the $\lambda$1400 feature falls outside the spectral range.
This leaves a sample of 22 objects in our VP06 sample.  Many of the
same objects are used by Park et al. (2013), however, they add 
PG 0804+761, Mrk 290, and NGC 4593, but exclude Mrk 110, Mrk 79, and NGC 4151.  
We also exclude NGC 3516, NGC 4051, and PG 1307+085
as we did from VP06 because of absorption or wavelength coverage issues
that make a peak flux measurement very uncertain.  
Our edited Park et al. (2013) sample has 23 objects.

Table 1 lists the spectra we measured and our adopted peak
$\lambda$1400/C~IV measurements for these two samples.  
We do not use these peak $\lambda$1400/C~IV measurements 
directly, only to select a well-matched sample with similar mean and 
dispersion from our SED sample for a self-consistent analysis. 
Table 1 includes peak $\lambda$1400/C~IV measurements
for a typical simulated RM sample matched to
Park et al. (2013).  Again, we want to analyze our 
own objects with our own consistent measurements and avoid 
any artificial biases associated with differences in measurement methodology.

We simulated a dozen RM samples, drawing 23 objects from the parent sample of
Paper 1, matching a quasar with log peak $\lambda$1400/C~IV to within 0.1 dex (25\%)
of a corresponding object from Park et al. (2013).  We selected one of these as 
our ``simulated RM sample'' on the basis of a very close match in log peak $\lambda$1400/C~IV
(both $-$0.76).  Our choice for illustrative RM sample also possessed 
the simulated sample average value of $a$ = 6.87 based on intercept fitting
described in $\S$ 3, and was thus representative of the simulated samples and
not an outlier. 

\begin{table*}
 \centering
 \begin{minipage}{140mm}
  \caption{Log peak $\lambda$1400/C~IV for Reverberation-Mapped Samples}
  \begin{tabular}{@{}lllll@{}}
  \hline

   Name     & peak $\lambda$1400/C~IV$^a$  & Source (Instrument) & Date & Samples  \\
\hline
3C 110 & 0.10 & HST (FOS) & 1995-03-16 & Sim \\
3C 120 & 0.13 &	IUE (SWP) & 1994-02-19,27; 1994-03-11 & P13, VP06 \\ 
3C 207 & 0.16 & HST (FOS) & 1991-12-04 & Sim \\
3C 263 & 0.18 & HST (FOS) & 1991-11-06 & Sim \\
3C 334 & 0.13 & HST (FOS) & 1991-09-07 & Sim \\
3C 390.3 & 0.17 & HST (FOS) & 1996-03-31 & P13, VP06 \\
4C 01.04 & 0.12 & HST (FOS) & 1994-09-11 & Sim \\
4C 12.40 & 0.16 & HST (FOS) & 1995-02-26 & Sim \\
4C 41.21 & 0.17 & HST (FOS) & 1992-12-09,10 & Sim \\
4C 49.22 & 0.15 & HST (FOS) & 1995-05-07 & Sim \\
4C 73.18 & 0.16 & HST (FOS) & 1994-09-07 & Sim \\
Ark 120 & 0.17 & HST (FOS) & 1995-07-29 & VP06, P13 \\
Fairall 9 & 0.21 & HST (FOS) & 1993-01-22 & P13, VP06 \\
IRAS F07546+3928 & 0.09 & HST (STIS) & 2000-01-28 & Sim \\
Mrk 110 & 0.16 & IUE (SWP) & 1988-02-28,29 & VP06 \\
Mrk 279	& 0.15 & HUT & 1995-03-05,11 & VP06 \\
Mrk 279 & 0.12 & HST (COS) & 2011-06-27 & P13 \\
Mrk 290	& 0.11 & HST (COS) & 2009-10-28 & P13 \\
Mrk 335	& 0.16 & HST (FOS) & 1994-12-16 & VP06 \\
Mrk 335 & 0.15 & HST (COS) & 2009-10-31; 2010-02-08 & P13 \\
Mrk 509 & 0.18 & HST (FOS) & 1992-06-21 & VP06 \\
Mrk 509 & 0.14 & HST (COS) & 2009-12-10,11 & P13 \\
Mrk 590	& 0.21 & IUE (SWP) & 1991 Jan 14 & P13, VP06 \\	
Mrk 79  & 0.17 & IUE (SWP)  & 1979-11-15 & VP06 \\
Mrk 817 & 0.19 & IUE (SWP) & 1981-11-06,08 & VP06 \\
Mrk 817 & 0.24 & HST (COS) & 2009-08-04; 2009-12-28 & P13 \\
NGC 3783 & 0.16 & HST (FOS) & 1992-07-27 & VP06 \\ 
NGC 3783 & 0.12 & HST (COS) & 2011-05-26 & P13 \\
NGC 4593 & 0.20 & HST (STIS) & 2002-06-23,24 & P13 \\ 
NGC 5548 & 0.11 & HST (FOS) & 1993-04-26 & VP06 \\
NGC 5548 & 0.09 & HST (COS) & 2011-6-16,17 & P13 \\
NGC 7469 & 0.17 & HST (FOS) & 1996-06-18 & VP06 \\
NGC 7469 & 0.10 & HST (COS) & 2010-10-16 & P13 \\
OS 562 & 0.19 & HST (FOS) & 1992-08-11 & Sim \\
PG 0026+129 & 0.16 & HST (FOS) & 1994-11-27 & VP06, P13 \\
PG 0052+251 & 0.17 & HST (FOS) & 1993-07-22 & VP06, P13 \\
PG 0804+761 & 0.31 & HST (COS) & 2010-06-12 & P13 \\
PG 0844+349 & 0.35 & HST (STIS) & 1999-10-21 & Sim \\
PG 0953+414 & 0.17 & HST (FOS) & 1991-06-18 & VP06, P13 \\
PG 1100+772 & 0.14 & HST (FOS) & 1993-02-03 & Sim \\
PG 1114+445 & 0.20 & HST (FOS) & 1996-05-13 & Sim\\
PG 1226+023$^b$ & 0.35 & HST (FOS) & 1991-01-14,15,16 & VP06, P13, Sim \\
PG 1229+204 & 0.23 & IUE (SWP) & 1982-05-01 & VP06, P13 \\
PG 1411+442 & 0.33 & HST (FOS) & 1992-10-03 & Sim \\
PG 1425+267 & 0.14 & HST (FOS) & 1996-06-29 & Sim \\
PG 1426+015 & 0.16 & IUE (SWP) & 1985-03-01,02 & VP06, P13 \\
PG 1440+356 & 0.30 & HST (FOS) & 1996-12-05 & Sim \\
PG 1512+370 & 0.09 & HST (FOS) & 1992-01-26 & Sim \\
PG 1613+658 & 0.33 & IUE (SWP) & 1991-02-25 & VP06 \\ 
PG 1613+658 & 0.26 & HST (COS) & 2010-04-08,09,10 & P13 \\
PG 1626+554 & 0.23 & HST (FOS) & 1996-11-19 & Sim \\
PG 2130+099 & 0.23 & HST (HRS) & 1995-07-24 & VP06 \\
PG 2130+099 & 0.20 & HST (COS) & 2010-10-28 & P13 \\ 
PG 2214+139 & 0.21 & HST (STIS) & 2000-06-19 & Sim \\ 
PG 2251+113 & 0.17 & HST (FOS) & 1991-10-23 & Sim \\ 
PKS 2216$-$03 & 0.25 & HST (FOS) & 1992-08-27,28 & Sim \\
\hline
\end{tabular}
$^a$\ As discussed in the text, fitting errors range up to about $15\%$
(see also Runnoe et al. 2013), but measurements made at different epochs 
can differ more than this, reflecting a larger intrinsic scatter.

$^b$\ The spectrum used by VP06 and P13 does not include the $\lambda$1400 feature, so we have used the measurement from Runnoe et al. (2013) for all samples.
  
\end{minipage}
\end{table*}

Figure 3 displays histograms of log (peak $\lambda$1400/C~IV) for 
the samples of Paper 1, the PGX subsample (Shang et al. 2007),
our simulated RM sample, and the VP06 and Park et al. (2013)
RM samples edited as described above.  
Log distributions of this quantity, like some other
EV1 properties such as the line ratio of [O III]/Fe II, appear
more normally distributed than the linear values.

We also note here that the peak $\lambda$1400/C~IV measurements from the 
RM samples do not correlate with the differences between the 
C~IV and H$\beta$ line widths as they do for our sample as seen in Paper I.  
The lack of correlation is not inconsistent with our previous findings,
but rather reflects the relatively narrow range in peak $\lambda$1400/C~IV
values.  If the RM samples were unbiased, they would be expected to 
have mean values and distributions more closely resembling that 
of the PGX sample, which shows a wider range of peak $\lambda$1400/C~IV
values shifted to larger values in the histograms of Figure 3.

\section{Analysis}

Using our data from Paper 1, we first demonstrate quantitatively 
how changing the EV1 distribution systematically changes the intercept
of the basic C~IV black hole scaling relationships.
Second, we compare the intercepts obtained for our nearly complete
PGX subsample and our simulated RM subsample to determine the 
approximate size of the EV1 bias in black hole masses based on 
C~IV scaling relationships from VP06 and Park et al. (2013). 
We conduct our analysis using both FWHM and $\sigma_{line}$.


\subsection{Ultraviolet EV1 and the Intercept Bias}

We first explore the effect of a bias in EV1, using different 
samples with systematically different peak ($\lambda$1400/C~IV),
on the intercept of the C~IV scaling relationship.  To do this 
we employ the formalism of VP06 with minor modifications.
We compute the unscaled mass $\mu$ from equation 3,
explicitly setting the luminosity exponent $\gamma = 0.5$.
Since our goal is to focus only on the effect 
of EV1 distributions on mass offsets, the problem is simplified if we can
fix as many parameters
as possible, and there is currently no evidence that the radius-luminosity 
relationship depends on EV1.
We use both FWHM$_{C~IV}$ or $\sigma_{C IV}$\ in place of $\Delta V$ 
in separate fits.

Figure 4 plots black hole mass as determined by the H$\beta$\ scaling 
relationship from VP06 on the Y-axis, and $\mu$ on the X-axis for both measures
of $\Delta V$ using the same data and measurements as Paper 1.  
Symbols are coded for different ranges of peak ($\lambda$1400/C~IV) as 
indicated, and a typical error bar is shown for an H$\beta$ mass uncertainty
of 0.4 dex (VP06) and 0.06 dex for peak ($\lambda$1400/C~IV).
Strong correlations are present, consistent with Paper 1, VP06, and Park
et al. (2013).  
Furthermore it is clear that the different subsamples are 
segregated, which we can quantify by determining the required intercepts to add to the 
unscaled mass to predict the H$\beta$-based mass.

\begin{figure*}
\begin{minipage}[!b]{8cm}
\centering
\includegraphics[width=8cm]{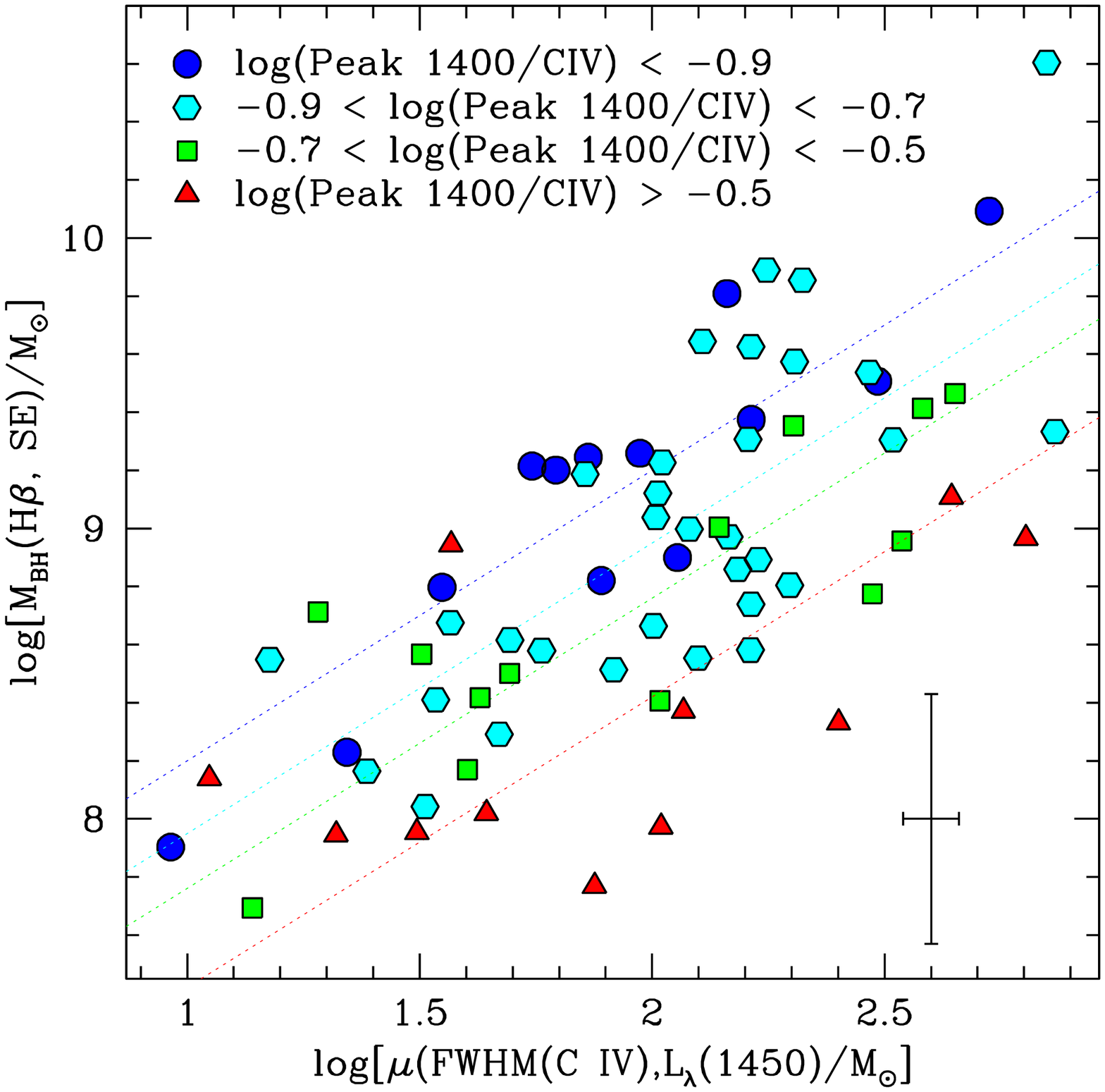}
\end{minipage}\hspace{0.6cm}
\hspace{0.6cm}
\begin{minipage}[!b]{8cm}
\centering
\includegraphics[width=8cm]{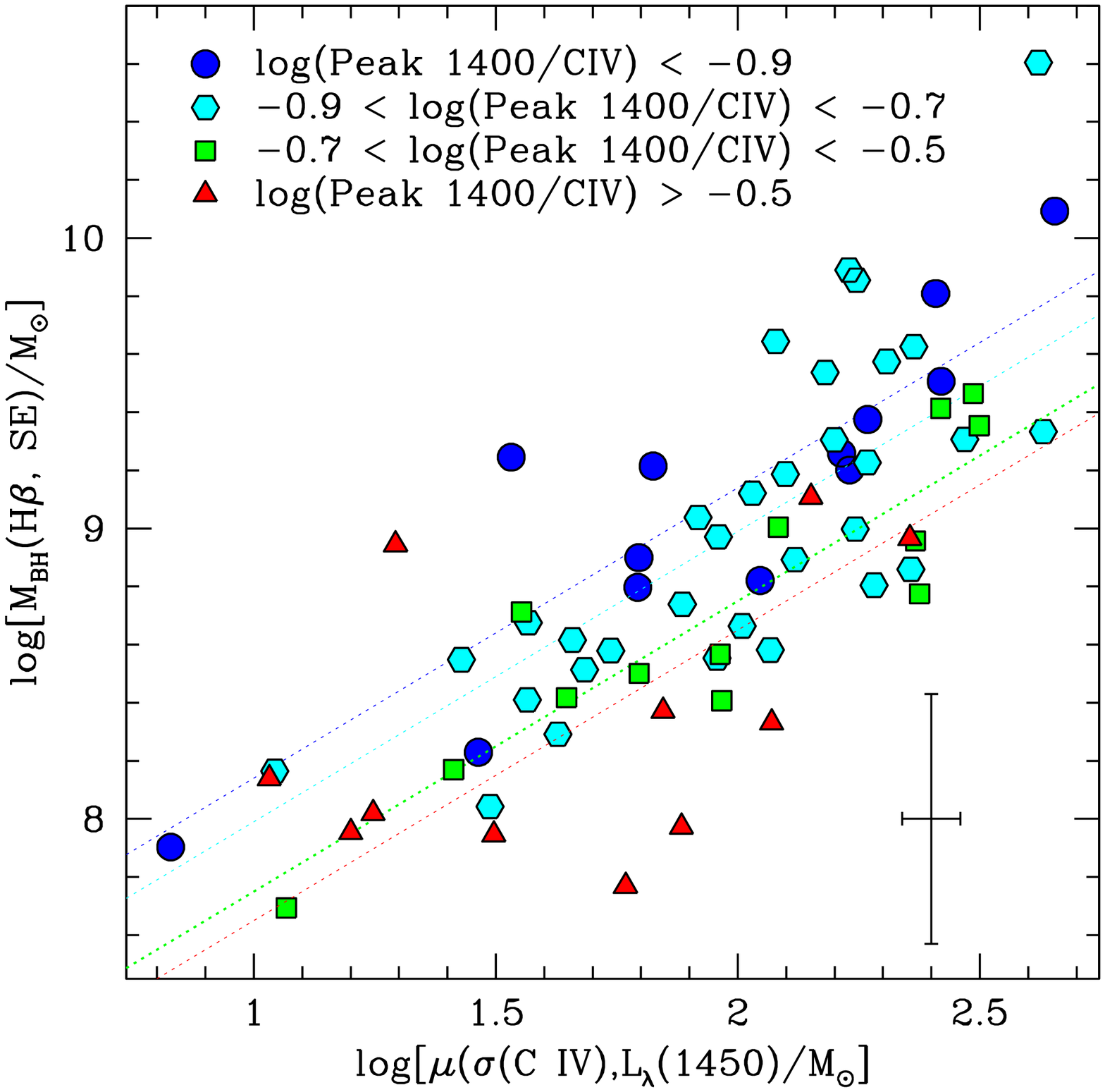}
\end{minipage}
\hspace{0.6cm}
\caption{H$\beta$-based single-epoch black hole masses calculated using 
the equation of VP06 (and tabulated by Tang et al. 2012) 
versus FWHM-based C~IV virial products labeled $\mu$ (left panel) for 
the sample of Runnoe et al. (2013a).  Symbols are color-coded and differently
shaped according to their value of log peak $\lambda$1400/C~IV as in Figure 3.  
A conservative error bar is shown in the lower right of the figure.  
The right panel
shows the same plot, except using a $\sigma_{line}$-based virial product.  
}
\label{fig:vp}
\end{figure*}


We used the robust least-squares fitting program GaussFit 
(Jefferys et al. 2013) to find a line minimizing the differences between
the two sides of equation 4, allowing only the zero-point
offset $a$\ to vary.  
We assumed typical uncertainties for all quantities, which
includes $M_{\rm BH}$ estimates (0.43 dex, e.g. VP06) 
and on parameters used to compute $\mu$ (10$\%$ on 
$\Delta$V and 3\% on continuum luminosity -- see Paper 1 and Shang et al. 
2011).  The fitting procedure provided the optimal offset $a$ and a 
corresponding 1 $\sigma$ uncertainty.  The fitting results
are given in Table 2 and show a strong and systematic trend: as the 
sample mean of peak ($\lambda$1400/C~IV) increases, the intercept $a$ 
decreases when using either FWHM or $\sigma_{line}$ of C~IV.
Again using GaussFit, we computed a best-fit line
to determine these relationships, for which we have assumed the 
standard error in the mean for the uncertainty on the 
log peak ($\lambda$1400/C~IV) term for each subsample, 
and the tabulated uncertainties on $a$.  
These fits are shown in Figure 5 and can be written
for FWHM-based C~IV masses:

\begin{equation}
$$a = (-1.18 \pm 0.05)( {\rm log\  peak}\ \lambda 1400/ {\rm C~IV}) + (6.01 \pm 0.04)$$
\end{equation}

and for $\sigma_{line}$-based C~IV masses:

\begin{equation}
$$a = (-0.80 \pm 0.12) ( {\rm log\ peak}\ \lambda 1400/ {\rm C~IV}) + (6.33 \pm 0.09)$$
\end{equation}

We note that the agreement in the slope reported in Paper 1, 
in the case of the FWHM-based masses, is good, $-1.18\pm0.05$ 
compared with $-1.227\pm0.136$ (their eq. 3).  
In the case of $\sigma_{line}$-based C~IV masses, the slope is flatter but
there is less consistency, $-0.80\pm0.12$ here compared to $-0.220\pm0.068$ 
(their eq. 4).  As before, we conclude this ultraviolet EV1 correction works well
for FWHM-based scaling relationships, but also may be weaker when applied
to $\sigma_{line}$-based scaling relationships.  As discussed in Paper 1,
this may be the result of FWHM being more sensitive to the EV1 variation,
in the form of a non-reverberating C~IV ILR component, than $\sigma_{line}$.

\begin{figure*}
\begin{minipage}[!b]{8cm}
\centering
\includegraphics[width=8cm]{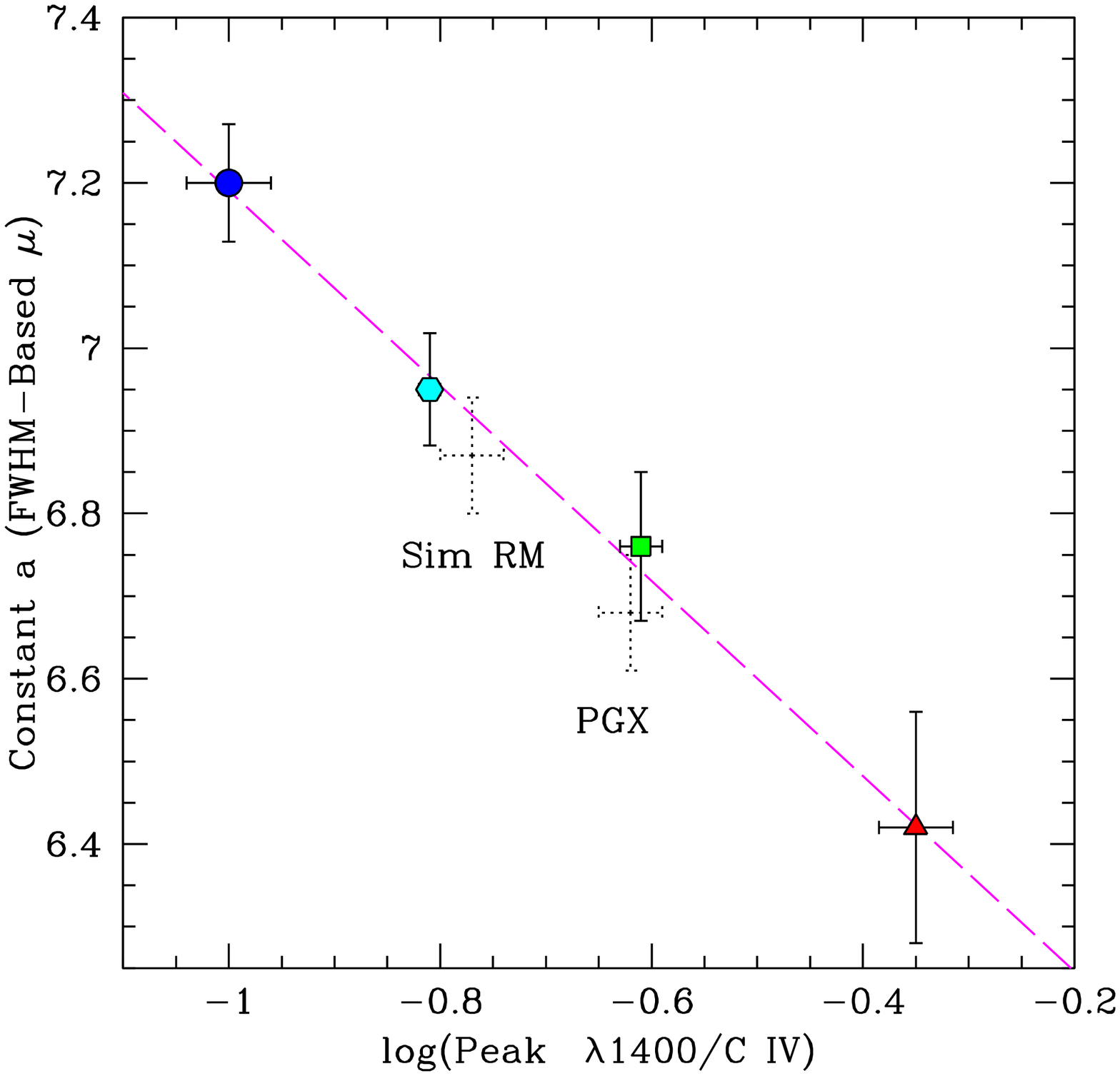}
\end{minipage}\hspace{0.6cm}
\hspace{0.6cm}
\begin{minipage}[!b]{8cm}
\centering
\includegraphics[width=8cm]{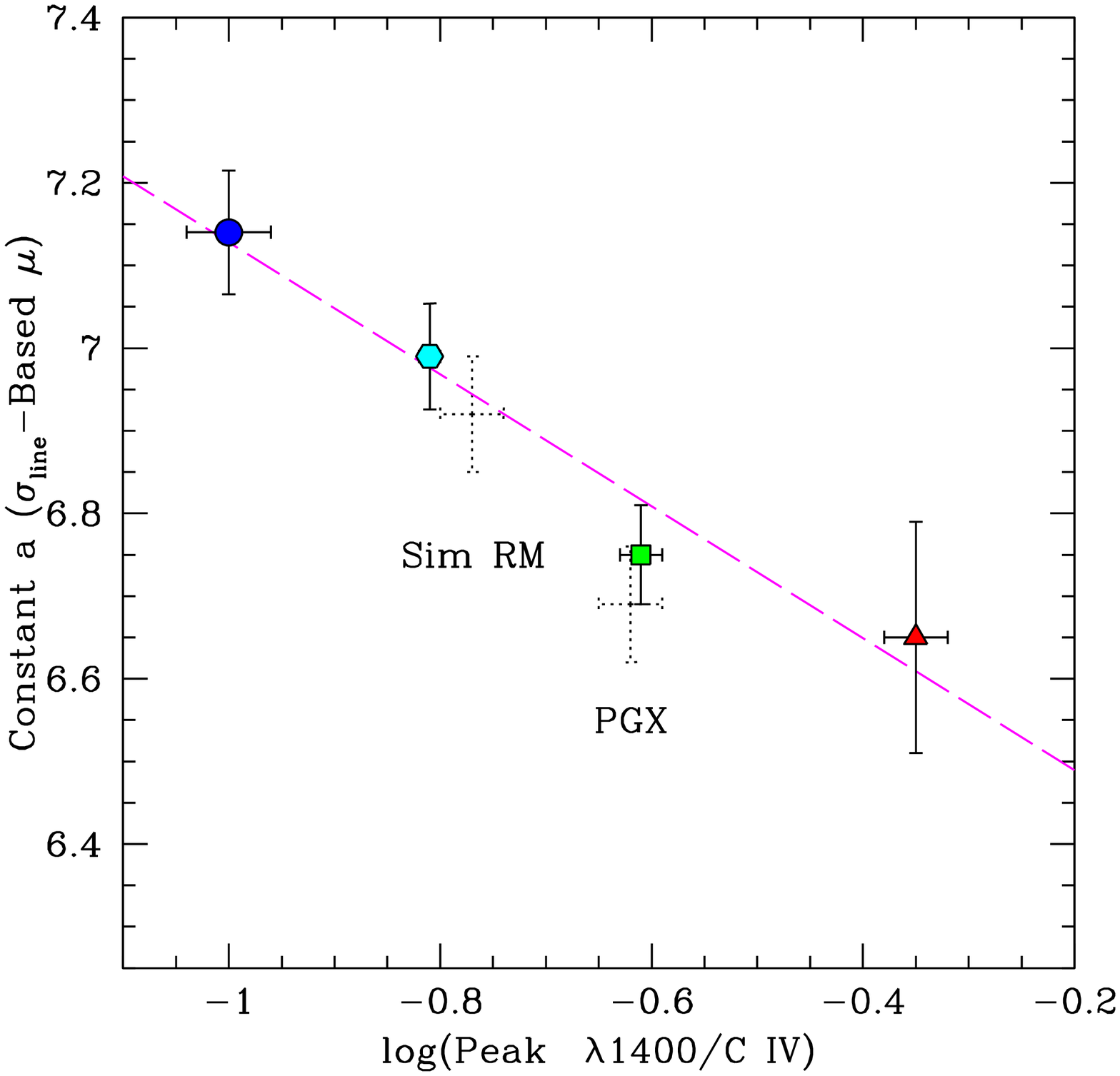}
\end{minipage}
\hspace{0.6cm}
\caption{
The fitted-line intercept $a$\ versus log peak $\lambda$1400/C~IV for the 
four binned samples, with symbols and colors to match previous figures,
for the FWHM-based (left panel) and $\sigma_{line}$-based virial products.
The errorbars displayed represent 1 $\sigma$ errors on a and the 
standard error of the mean for the log(peak $\lambda$1400/C IV) subsamples.
We note the typical fitting uncertainties on $a$\ quoted by VP06 are less 
than 0.02 dex, much smaller than the effect illustrated here.
}\label{fig:linefits}
\end{figure*}

\begin{table*}
 \centering
 \begin{minipage}{140mm}
  \caption{C~IV Mass Scaling Relationships: EV1 Subsamples and Zero Points}
  \begin{tabular}{@{}ccccc@{}}
  \hline
log peak $\lambda$1400/C~IV Range & $<$log Peak $\lambda$1400/C~IV$>$ $\pm$ SEMean & N & FWHM Zero Point $a$ & $\sigma_{line}$ Zero Point $a$ \\
\hline
$-0.90$ to $-1.10$ & $-1.00 \pm 0.04$ & 12 & $7.20 \pm 0.07$ & $7.14 \pm 0.08$ \\
$-0.70$ to $-0.90$ & $-0.81 \pm 0.01$ & 32 & $6.95 \pm 0.07$ & $6.95 \pm 0.07$ \\
$-0.50$ to $-0.70$ & $-0.61 \pm 0.02$  & 13 &  $6.76 \pm 0.09$ & $6.75 \pm 0.06$ \\
$-0.10$ to $-0.50$ & $-0.35 \pm 0.04$  & 11 &  $6.42 \pm 0.14$ & $6.65 \pm 0.14$ \\
\hline
\end{tabular}
\end{minipage}
\end{table*}

\subsection{PGX and Reverberation-Mapped Samples}

We used the above equations along with the mean values of 
log peak $\lambda$1400/C~IV for the VP06 and Park et al. (2013) 
RM samples to estimate their zero-point offset
and compare them to those of the nearly complete PGX subsample in order to
compute the bias.  Perhaps better, however, is to fit $a$\
directly as in equation 4 for our simulated RM sample. In any case,
both approaches give similar results.  See Table 3.

\begin{table*}
 \centering
 \begin{minipage}{140mm}
  \caption{Zero Points for PGX and Simulated RM Samples}
  \begin{tabular}{@{}ccccc@{}}
  \hline
Sample & $<$log peak $\lambda$1400/C~IV$>$ $\pm$ SEMean& N & FWHM Zero Point $a$ & $\sigma_{line}$ Zero Point $a$ \\
\hline
Simulated RM &  $-0.76 \pm 0.03$ & 23 & $6.87\pm0.07$ & $6.92\pm0.07$ \\
PGX  & $ -0.62\pm0.05 $ & 22 & $6.68\pm0.10$ & $6.69\pm0.07$ \\
\hline
\end{tabular}
\end{minipage}
\end{table*}

The mean value of log peak $\lambda$1400/C~IV is $-$0.62 in the 
PGX sample, and $-0.76$\ with a tighter distribution 
in the simulated RM sample.  This latter value 
is consistent with the average value of $-$0.76 computed for
the values of our edited Park et al. (2013) sample.

Taking into account that our PGX subsample is not totally complete and
only 22 objects, and that we have had to edit the VP06 and Park et al. (2013)
samples slightly, which are each only 23 objects, and that matching 
those samples can only be done approximately, we prefer to be 
quantitatively conservative. The zero-point offset $a$\ in the 
FWHM-based C~IV black hole scaling relationship derived using 
RM samples differs from that obtained for the PGX sample,
and will yield masses that are
systematically  high by almost $\sim$0.2 dex or $\sim$50\%.
There is a similar bias for $\sigma_{line}$-based C~IV masses, which 
is smaller but on the same order.

\section{Discussion}

We are now entering an era of identifying first and second-order 
corrections to SE quasar black hole scaling relationships.  Real
line profiles are complex and do not seem to always represent 
gas moving in a purely virial manner.  Moreover, accuracy and 
precision are both important, and efforts must be made not only
to reduce scatter in scaling relationships, but also to get the 
zero point right.  Specifically, which measurements should be 
made to yield the best black hole masses? 

For instance, is the choice of measuring $\Delta$V using FWHM or 
$\sigma_{line}$ preferred?  A number of articles have suggested 
that the latter may be a better choice (e.g. Collin et al. 2006).
We have shown here $\sigma_{line}$-based masses also suffer an EV1 bias.
The $\sigma_{line}$-based masses can be also biased when measured using spectra
with low signal-to-noise (SNR) ratios (e.g. Denney et al 2013).  
Thus FWHM, being a more robust measurement for typical lower SNR 
quasar spectra from the Sloan Digital Sky Survey (SDSS,
e.g., Shen et al. 2011) is likely preferred in that case.  Working on FWHM-based scaling 
relationships is therefore valuable even if $\sigma_{line}$ may be a less
biased measurement in high-SNR spectra.



\subsection{Current and Future Bias Corrections}

If an investigator wants the best black hole mass with only rest-frame
ultraviolet spectra of low or moderate SNR, what is the best course to take?

If the object is in the SDSS Data Release 7, 
they can adopt the Shen et al. (2012)
values already given to them, based on the VP06 prescription,
but increase the value by 0.2 dex or 50\% to take into account the
typical EV1 bias.  If they want to improve on that average EV1 correction, 
they can make additional measurements
of the peak heights of the $\lambda$1400 feature and C~IV and apply 
equation 3 of Paper 1.  This is also a reasonable course for updating
any C~IV based masses already in use that were computed using VP06 formulations.

This is probably not ideal, however, given the changes between the work
of VP06 and that of Park et al. (2013), 
who made many small improvements that together
led to significantly different C~IV scaling relationships.  An average
correction for EV1 bias can still be applied, increasing masses by $\sim$50\%.
If measurements of the peaks of $\lambda$1400 feature and C~IV can be
made, then a better, individualized correction is possible:

\begin{equation}
M_{\rm BH} = M_{\rm BH}({\rm FWHM_{C~IV}, P13}) - 1.23\ {\rm log\ peak}\ \frac{\lambda 1400}{\rm C IV} - 0.91
 \end{equation}

The above equation assumes the Paper 1 slope on the EV1 term and a 
RM sample mean of log peak $\lambda$1400/C~IV of $-$0.76.
While we have provided a second decimal place on the numbers in the equation, 
we recommend rounding results to only the first decimal place. 
We suggest the Paper 1 slope ($-1.23$ rather than $-1.18$) because
the previous analysis does not bin data which can 
have a small effect on the fitted slope, and note that the larger
uncertainty is the result of conservative errors on individual points compared to the actual
scatter in the data points that we used for the subsamples.
This equation also assumes that our EV1 correction does not change for 
higher luminosity, higher redshift quasars, but this should be explicitly and 
quantitatively investigated.
The correction could change if the ``Baldwin effect'' (Baldwin 1977),
the inverse correlation between emission line equivalent width and 
continuum luminosity, differs between C~IV and the $\lambda$1400 feature,
or if it differs between the contaminating ILR component and the rest
of the C~IV emission line.  
Shang et al. (2003) found that the narrower line core of C~IV correlated
with luminosity in the sense of the Baldwin effect, and therefore 
a dependence on luminosity seems plausible, if not likely.
Furthermore, at higher redshifts whatever unknown
property or properties driving EV1 could differ from those in our
lower redshift samples, or differ in their manifestation.

\subsection{Implications for Current and Future RM Campaigns}

It is not necessarily necessary to give up using RM samples to 
derive black hole mass scaling relationships.  In principle such samples
have the most reliable black hole masses and it is desirable to use them
when possible.
Ideally, these samples can be expanded to include a larger range in EV1, 
particularly objects at the narrow-line Seyfert 1 (NLS1) end of the trend 
at high values of peak $\lambda$1400/C~IV, that are currently underrepresented,
at least in the case of also having rest-frame ultraviolet spectra
covering the C~IV emission line.  
New ultraviolet {\em HST} spectroscopy of NLS1s that have been 
reverberation mapped (e.g., Du et al. 2014) would be quite useful
in expanding the sample of RM quasars with a wider range of EV1 properties.
RM campaigns should, more generally, endeavour to include the full
range of all types of broad-lined AGNs in order to avoid systemic biases
of all types, and not just target the easiest and most cooperative
objects.

We must be concerned about some NLS1s, however, given suspicions that
the presence of extreme high-ionization winds emitting C~IV might 
prevent accurate mass estimation (e.g., Vestergaard 2011).  Additionally,
the currently existing EV1 bias could result from a physical effect,
that NLS1s may not vary as strongly as other broad-lined Seyfert Galaxies 
and quasars (e.g., Ai et al. 2013), making
them more challenging targets for reverberation campaigns.
Another potential issue is that EV1 properties may result in biases of
not only the C~IV derived masses, but of H$\beta$ derived masses.
Du et al. (2014) find that two of their NLS1s with extremely high accretion 
rates have time lags significantly shorter than expected based on existing
R$_{\rm BLR}$ - L relationships, which may lead to systematic overestimation
of black hole masses for similar objects using existing H$\beta$ scaling 
relationships.  A statistically larger number of reverberation-mapped 
AGNs is required to more fully investigate this result.

Despite these caveats, there is reason for optimism as improvements 
are clear and more can be expected in the future given these and 
other approaches.  Of particular interest is the new, large 
reverberation mapping effort described by Shen et al. (2015), 
which promises to deliver dozens of new broad-line time lags
for a homogeneously selected sample of quasars likely including
the full range in EV1.
Quasar black hole mass determinations will improve, and RM efforts 
are still needed to drive improvements.
 
\section{Summary}

We have explicitly demonstrated how samples biased in UV probes EV1 produce 
biased black hole mass scaling relationships.  We have also determined 
quantitatively the size of the effect and its general consistent with our
previous work.  We have shown how to make both average and individualized
corrections to black hole masses estimated using the C~IV line based
on equations derived from reverberation-mapped samples, which are popular
but heterogeneous and biased in EV1.  They, on average, lead to mass 
estimates that are $\sim$0.2 dex or about 50\% too high.

\section*{Acknowledgments}

We would like to thank Kelly Denney for helpful discussions during the 
preparation of this work.  M. S. Brotherton acknowledges 
support through the Space Telescope Science Institute, AURA
through Grant HST-AR-13237.01-A.
Z. Shang acknowledges support by the National Natural Science Foundation 
of China through Grant No. 10773006 and Tianjin Distinguished Professor Funds.

\label{lastpage}

\end{document}